# Mass Conservation and Inference of Metabolic Networks from High-Throughput Mass Spectrometry Data

Pradeep Bandaru[1,2], Mukesh Bansal[1], and Ilya Nemenman[3*]

[1]Center for Computational Biology and Bioinformatics, Joint Centers for Systems Biology, and Columbia Initiative in Systems Biology, [2]Department of Biomedical Informatics, Columbia University, New York, NY 10032, USA

[3]Departments of Physics and Biology and Computational and Life Sciences Strategic Initiative, Emory University, Atlanta, GA 30322

[*]Contact: ilya.nemenman@emory.edu

## Abstract

We present a step towards the metabolome-wide computational inference of cellular metabolic reaction networks from metabolic profiling data, such as mass spectrometry. The reconstruction is based on identification of irreducible statistical interactions among the metabolite activities using the ARACNE reverse-engineering algorithm and on constraining possible metabolic transformations to satisfy the conservation of mass. The resulting algorithms are validated on synthetic data from an abridged computational model *of Escherichia* coli metabolism. Precision rates upwards of 50% are routinely observed for identification of full metabolic reactions, and recalls upwards of 20% are also seen.

.

## Introduction

Prior to the widespread availability of annotated metabolic databases, metabolic network reconstruction was carried out primarily with biochemical assays of enzymatic activity (*1-3*), resulting in a pathway-centric depiction of chemical reactions occurring in a cell. For some organisms, it has been possible to assemble these data into genome-wide metabolic networks by means of various Metabolic Flux Analysis (MFA) techniques (*4-9*). Now, recent developments in the burgeoning field of metabolic profiling (*10-12*) and especially mass-spectrometry metabolomics (*13-17*), which aims at high-throughput, real-time characterization of the entire cellular metabolic state, have opened up yet another way of approaching the metabolic network reconstruction problem, focusing on statistical interactions among metabolites. This parallels the transition that had happened in the analysis of transcriptional regulatory networks with the advent of gene expression profiling (*18-22*), which similarly characterizes the genome-wide transcriptional state of the cell.

Specifically, distinct cellular phenotypes, phases of the cell cycle, or intrinsic and extrinsic perturbations result in changes in cellular metabolite concentrations. However, even with such changes, concentrations of metabolites that transform into each other should stay correlated, and the observed dependencies can be used to predict metabolic reactions computationally even if the identities of the metabolites are unknown, preventing the application of MFA methods. We attempted this approach (*23*) using the ARACNE statistical reverse engineering method, first developed in the context of inferring transcriptional networks from mRNA expression profiles (*19, 24, 25*). However, mass-spectrometry methods provide a wealth of information about the metabolites in addition to their abundances: MS-MS methods and isotopic labeling (*13-17*) can recover the molecular structure, and, especially crucial for this paper, masses of metabolites are typically measured to the accuracy of $10^{-4} \ldots 10^{-5}$. Since mass must be conserved in any metabolic transformation, this information presents an additional source of data for reverse engineering methods that has not been widely used. Namely, a statistical dependence among metabolites can indicate an actual metabolic transformation only if the putative chemical reaction constructed from these metabolites conserves mass. This rule can be applied even when identities of metabolites are unknown and, therefore, it is especially useful for high-throughput global metabolic profiling where spectral peaks cannot necessarily be identified.

In this paper, we present a *M*ass-*C*onstrained adaptation of the ARACNE algorithm, ARACNE-MC, which should be considered as a first foray into the field, laying the foundation for future studies. Note that, in the case of metabolism, we are not content with knowing just the statistical dependencies among the metabolites, even if they correspond to bona-fide metabolic transformations. Instead we aim at a substantially more complicated task of reconstructing complete metabolic reactions, identifying all of their substrates and products.

We test the algorithm on reduced toy models of the *Escherichia coli* metabolome with *in silico* simulated metabolic profiles. Applications to real-world biological problems will have to wait for larger experimental datasets.

# Results

## Outlines of the algorithms

### The ARACNE Algorithm

The basis of the reverse engineering approach we undertake is the notion that molecular species that participate in biochemical reactions have statistically dependent expressions (*23*). Within the ARACNE framework (*24, 25*), one views metabolite expressions as random variables sampled from stationary probability distributions. This randomness accounts for effects of unknown states of unobserved metabolites and other chemical species and of the experimental noise. Chemical transformations correspond to nonzero multivariate statistical dependencies among metabolite concentrations (*26*). In particular, the relevant measure of statistical dependency between two variables $(c_1, c_2)$ is their mutual information (MI) (*27*)

$$I[c_1; c_2] = \left\langle \log_2 \frac{P(c_1, c_2)}{P(c_1)P(c_2)} \right\rangle_{P(c_1, c_2)}, \tag{1}$$

where $P(c_1, c_2)$ denotes their joint probability distribution, $P(c_i)$ are the marginals, and $\langle \ldots \rangle_P$ is the average over the distribution $P$.

The MI is generally nonzero for bona fide interacting metabolites, but it may also be nonzero for chemicals that are connected through an intermediate and do not transform into each other directly. In fact such *false positives* are generally a bigger problem than *false negatives* (i.e., missing a true interaction) in computational networks reverse engineering: false positives are plentiful and lower the confidence in the validity of every specific prediction.

The ARACNE algorithm (*24, 25*) eliminates some of the false positives by using the data processing inequality (DPI) (*27*) to isolate statistical interactions that have the highest chance of corresponding to true biological transformations. Specifically, under certain assumptions that are often applicable in transcriptional (*24*) and metabolic (*23*) contexts, if

$$I[c_1; c_3] \leq \min\left(I[c_1; c_2], I[c_2; c_3]\right) \tag{2}$$

then the $c_1 \leftrightarrow c_3$ interaction is indirect. Hence ARACNE starts with a fully connected graph of the measured chemical species as putative interaction partners, compares MIs for every triplet of chemical species in the dataset, and removes the weakest pairwise interaction in every such triplet from further consideration. Practical complications in the application of ARACNE revolve around accurate, unbiased estimation of MI and of the threshold (5-15% for typical applications) above which a difference between two MI values becomes significant for the DPI application (*19, 24*). Though developed for reverse engineering transcriptional networks, ARACNE has also been validated in the context of synthetic metabolic networks (*23*).

**The ARACNE-MC Algorithm**

As we have emphasized, mass-spectrometry provides additional information about the metabolic state of a cell, namely, masses of the metabolites. This creates extra means for eliminating false positive metabolic transformations: even if statistical dependencies suggest an interaction, the implied putative chemical reaction may not conserve mass and hence be impossible. To use this additional constraint, we propose the ARACNE-MC algorithm (MC stands for *Mass Constrained*). Like the original ARACNE, ARACNE-MC aims at reducing false positives, potentially at the cost of increasing the false negatives.

For a list of metabolites $\mu_a$ with masses $m_a$ and metabolic activities in $i$'th spectrometer run $c_{ai}$, we start by building a list of putative metabolic reactions allowed by mass conservation, focusing on a limited set of template reactions that are allowed in the analysis. In this paper, we consider only three templates (a) 1x1, $\mu_{l_1} \leftrightarrow \mu_{r_2}$; (b) 1x2, $\mu_{l_1} \leftrightarrow \mu_{r_1} + \mu_{r_2}$; and (c) 2x2, $\mu_{l_1} + \mu_{l_2} \leftrightarrow \mu_{r_1} + \mu_{r_2}$ (indexes $l$ and $r$ stand for *left* and *right,* respectively). See Fig. 1 for example reactions of each template. Eliminating more complicated reactions from consideration will result in the elimination of bona fide statistical interactions, and hence in extra false negatives, which we accept. We build a list of all putative reactions that fall into the allowed template classes and satisfy the mass conservation, $\left|\sum m_{l_i} - \sum m_{r_i}\right| \le \varepsilon \sum m_{l_i}$, where $\varepsilon$ is the mass equality tolerance, set to $\varepsilon = 10^{-4}$ throughout this paper. We refer to such reactions as *conforming* reactions. Identifying them has a computational complexity of $O(M^{\alpha})$, where $\alpha$ is the maximum number of metabolites on either side of the template (2 in this paper).

While possible in principle, a conforming reaction may not exist in a real cell due to a multitude of factors. If present, it should result in statistical dependencies among its reactants and products. There will be up to one such dependence for a 1x1 reaction (two choose two), three for a 1x2 reaction (three choose two), and six for a 2x2 reaction (four choose two). Therefore, we prune the list of conforming reactions by identifying those that are supported by statistical dependencies. To do so, we apply ARACNE to metabolite activity profiles, $c_{ai}$. Specifically, we estimate the pairwise MIs $I\left[c_a, c_b\right]$ using the algorithms of (*19*) and then apply the DPI to the list of MIs to select the interactions that have the highest chance of being direct. Then the conforming reactions are ranked by how many of their pairwise member metabolite interactions are identified as direct by ARACNE. We expect that reactions that are conforming *and* supported statistically will have a high chance to be bona fide metabolic reactions. The ARACNE-MC1 algorithm thus requires selection of the threshold for the number of ARACNE-supported metabolite pairwise interactions, and identifies all of the conforming reactions passing the threshold as putative reactions.

We notice that the same metabolic statistical interaction may be a part of multiple reactions. One "strong" reaction may be responsible for much of the MI associated with a particular link, and thus for the corresponding interaction surviving the ARACNE DPI application. However, ARACNE-MC1 would count this interaction in support of every conforming reaction to which it belongs. To avoid this multiple counting, we sort all conforming reactions by their "strength" as measured by the cumulative mutual information in all of its interactions. We then ensure that an interaction is counted as supporting only the strongest of its associated reactions; this is the ARACNE-MC2 algorithm.

Flowcharts of both algorithms are illustrated in Fig 2.

The algorithms, implemented in MATLAB, are available at http://menem.com/~ilya/wiki/index.php/Bandaru_et_al,_2010. The performance of the algorithms will depend on a variety of choices, such as the DPI and mass comparison tolerances, MI estimation parameters, or thresholds for the number of interaction supports needed to proclaim a reaction as existing. Some of these choices are explored later in the paper. For others, which we believe may differ dramatically for synthetic and for experimental data, we leave the detailed analysis to future publications.

## Synthetic Tests of ARACNE-MC

### Data Generation and Performance Metrics

To validate performance of ARACNE-MC, we used the Kyoto Encyclopedia of Genes and Genomes (KEGG) *(31)* to create synthetic metabolic networks, which then served as a source of simulated data for tests. KEGG provides a detailed description of metabolites and reactions found in the metabolic pathways of various model organisms. The entirety of the metabolic pathways of *Escherichia coli* were downloaded and pruned to include only mass-balanced reactions of the three types shown in Figure 1. Two different synthetic networks were constructed to test the performance of ARACNE-MC on metabolic networks of different sizes: a *small* network, containing 86 unique metabolites and 50 metabolic reactions, and a *large* network with 218 metabolites and 136 reactions. Detailed specifications of the networks are available at http://menem.com/~ilya/wiki/index.php/Publications/Bandaru_et_al.,_2010.

Since a majority of kinetic rates are unknown in KEGG, for each of the reactions we randomly generated forward and backward rates, with the forward to backward ratio being, on average, a hundred. We used the open-source COPASI software (*28*) to simulate the dynamics of the network multiple times, perturbing the kinetic rates each time by a random multiplicative factor with a standard deviation of 15%. This is similar to the approach in (*24*) and is supposed to represent changes to the rates due to different extracellular environments and phenotypic and metabolic states of the cells. Each simulation started with equal metabolite concentrations of 1uM and was run to a steady state. One thousand steady states were simulated this way and used as synthetic metabolic profiles for input to ARACNE-MC.

To measure the algorithm performance, we choose the metrics of *precision* and *recall*., Precision, $\pi = N_{TP} / \left(N_{TP} + N_{FP}\right)$, measures the fraction of true predictions among all predictions (where indices $T$, $F$, $P$, and $N$ stand for *true, false, positives,* and *negatives*). Precision corresponds to the expected success rate in the experimental validation of computational predictions. Similarly, recall, $\rho = N_{TP} / \left(N_{TP} + N_{FN}\right)$, indicates the fraction of all reactions recovered by the algorithm. Precision and recall values of 1 indicate perfect performance. However, there is generally a tradeoff between the two. We emphasize that all of these metrics are calculated based on recovery of complete metabolic reactions, rather than simple statistical correlations among the metabolites, as in (*23*).

**ARACNE-MC Performance**

We performed two tests to verify that accurate knowledge of both the metabolite profiles and the mass constraints is necessary for ARACNE-MC1 and 2 performance. Firstly, we randomized entries in the mass-conforming reactions while leaving the statistical relations intact. Secondly, we randomized metabolic profiles while preserving the correct mass constraints. In both cases, ARACNE-MC1 and 2 failed to predict metabolic reactions beyond chance, indicating an equal reliance on the two types of data (results not shown).

Table 1 details the results from the ARACNE-MC1 algorithm in the reconstruction of the large (218 metabolites) and small (86 metabolites) synthetic networks. Precisions well over 50% with significant recall rates are seen for many parameter combinations. The results suggest that the ARACNE algorithm is robust to changes in network size and topology. We emphasize again that, for our performance metrics, *all* substrates and products of a reaction must be predicted correctly for the reaction to be counted as correct. This is a very stringent performance test.

Performance of ARACNE-MC2 is illustrated in Table 2. We see that removing double counting of interactions has an effect of increasing the precision (sometimes to the maximum level of 1) with only a marginal loss in the recall.

To illustrate the dependence of the ARACNE-MC reconstruction on the DPI tolerance parameter, Table 3 shows performance of ARACNE-MC1 for the tolerance of 1 (i.e., no DPI applied). While performance is weaker compared to the tolerance of 0, the effect is not dramatic, suggesting weak sensitivity to the parameters. The specific best value of the parameter will likely depend on the size of the dataset and on the experimental noise, and will need to be established for each particular application independently as in (*29*)

Finally, a crucial feature of any computational reverse engineering algorithm is the dependence of its performance on the size of the experimental dataset. We test this for ARACNE-MC2 in Table 4. Specifically, both the precision and the recall degrade gracefully as the data set size decreases from 1000 to 100, and no meaningful reconstruction is possible when the size becomes comparable to the number of the analyzed metabolic species. This explains, in particular, why our application of the algorithms to the existing experimental data set of Ishii, et al. (*17*), which includes approximately 30 steady-state metabolic profiles *of Escherichia coli* and 195 metabolites in each profile, has failed.

## Discussion

The ARACNE-MC1 and 2 algorithms represent the first computational step towards identification of metabolic reaction networks from high-throughput mass-spectrometry profile data, armed with detailed knowledge of metabolite masses. Performance of the algorithm on synthetic data sets is encouraging, warranting further development and application to real-life data sets, when available. Selection of optimal values of many parameters of the algorithm, which we expect to depend on the details of the experimental data, will need to be performed at that time. Further, depending on the experimental resolution for many small, common metabolites, additional modifications of ARACNE-MC will need to be considered. In particular, to reduce the rate of false negatives, frequent interactions among common substrates (ATP, water, NADP, etc.) can be treated as supported statistically for every conforming reaction.

As implemented now, the algorithm is data-intensive, requiring more metabolic profiles than the number of considered metabolites. Current absence of such large datasets is the biggest obstacle in application of the algorithm to real-world problems. However, we expect that ion-mobility mass spectrometry with nanoliter chemostat cultures (*30*) will be able to provide the necessary amounts of data in the immediate future.

# Methods

### Synthetic Networks Generation

Synthetic metabolic networks were created from the KEGG database. Using *Escherichia coli* as a model system for these synthetic networks, we downloaded mass-balanced reactions randomly, so that the final analyzed network is representative of the metabolism of *E. coli.* A small synthetic network containing 86 unique metabolites and 50 metabolic reactions and a large synthetic network containing 218 metabolites and 136 metabolic reactions were generated. See the appendices for detailed descriptions of the metabolites in each network and their corresponding masses.

### Analysis and Simulation Parameters

Two main parameters of ARACNE algorithm are the p-value for accepting an MI estimate as nonzero and the DPI tolerance threshold. For the purposes of this study, the DPI tolerance was varied between 1 (no DPI application) and 0 (stringent edge elimination), and the p-value threshold was set to the default level of 1e-4. Additionally, the mass comparison relative tolerance was 1e-4.

# Acknowledgements

We especially thank Andrea Califano, who has been an invaluable help in the early stages of this research. We also thank William Hlavacek, Fangping Mu, Joel Berendzen, and Manjunath Kustagi for important contributions. PB and IN were partially supported by NIH/NIGMS under 1R21GM080216.

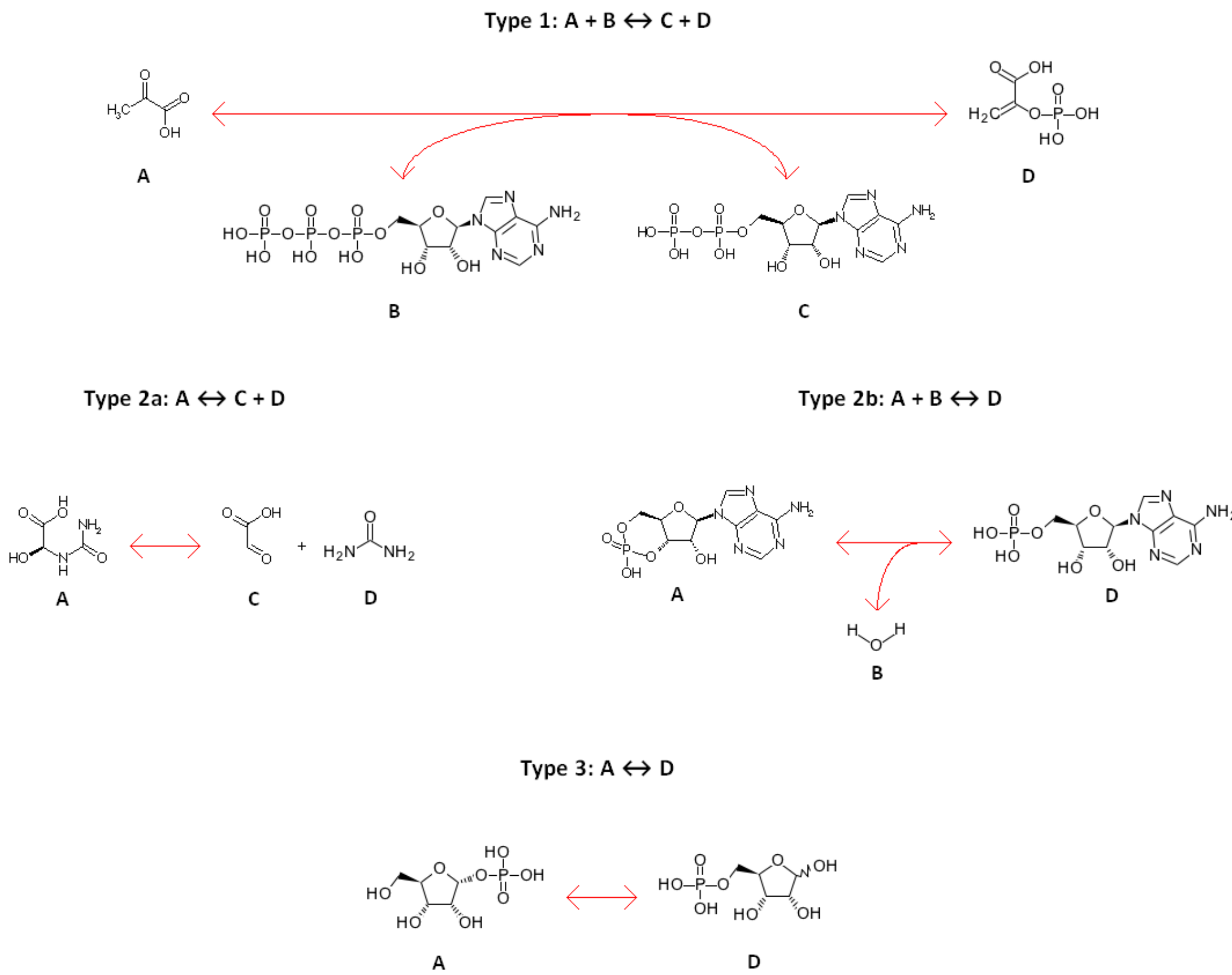

Figure 1. Examples of three types of metabolic reactions included in the synthetic network and generated by the mass constraints algorithm. Type 1 reactions are termed two-by-two reactions, Type 2a and 2b reactions are termed one-by-two reactions, and Type 3 reactions are termed one-by-one reactions. Chemical structures from KEGG are provided for illustration.

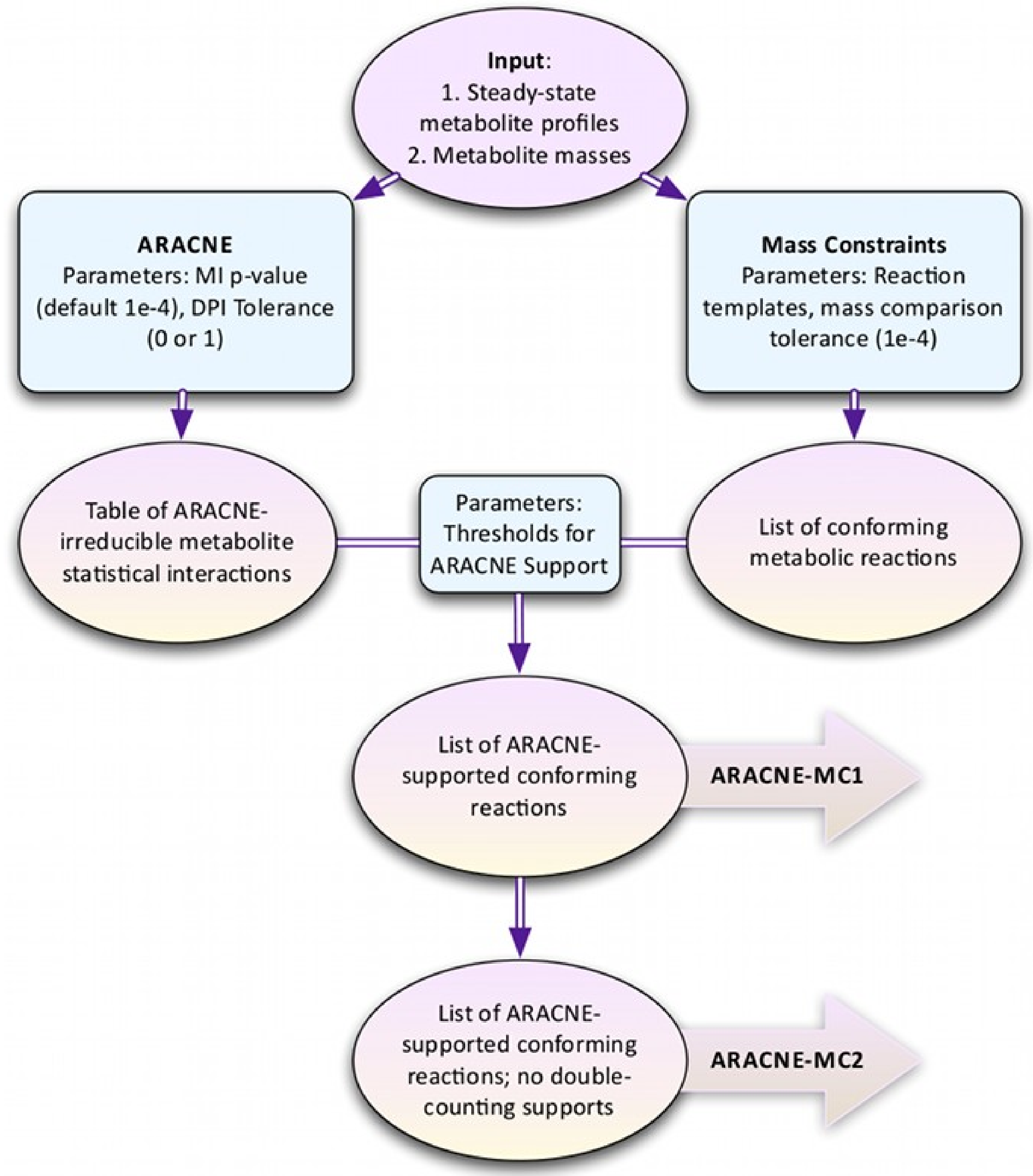

Figure 2. Flowchart of the ARACNE-MC algorithm.

| Template | ARACNE Interactions | TP | FP | Precision | Recall |
|---|---|---|---|---|---|
| 1x1 | 1 | 1/0/15 | 4/0/11 | 0.20/0/0.58 | 0.50/0/0.71 |
| 2x1 | 3 | 2/7/6 | 1/8/10 | 0.67/0.47/0.38 | 0.11/0.37/0.16 |
| | 2 | 6/14/20 | 2/17/33 | 0.75/0.45/0.38 | 0.32/0.74/0.53 |
| 2x2 | 6 | 0/5/9 | 0/21/2 | 0/0.19/0.82 | 0/0.17/0.12 |
| | 5 | 1/10/14 | 3/31/2 | 0.25/0.24/0.88 | 0.03/0.34/0.18 |
| | 4 | 4/18/20 | 6/98/29 | 0.40/0.16/0.41 | 0.14/0.62/0.26 |
| | 3 | 12/23/48 | 28/369/355 | 0.30/0.06/0.12 | 0.41/0.79/0.62 |
| | 2 | 21/28/61 | 227/969/77 | 0.08/0.03/0.02 | 0.72/0.97/0.79 |

Table 1: Performance of ARACNE-MC1 for different networks and DPI tolerances. The first value in each cell in the four right columns corresponds to the small networks with zero tolerance; the second value is for the small network with the tolerance of one, and the third value is for the large network. All data is for 1000 simulated metabolic profiles. When the DPI tolerance is 0, precisions over 80% are possible with recalls in the teens when large ARACNE support for a conforming reaction is requested, and recall of 25-50% still leaves the precision around 40%.

| Template | ARACNE Interactions | Surviving ARACNE Interactions | TP | FP | Precision | Recall |
|---|---|---|---|---|---|---|
| 1x1 | 1 | N/A | 0/0 | 0/0 | 0/0 | 0/0 |
| 2x1 | 3 | 3 | 1/3 | 0/0 | 1/1 | 0/05/0/08 |
| | | 2 | 2/6 | 0/0 | 1/1 | 0.11/0.16 |
| | | 1 | 2/6 | 0/0 | 1/1 | 0.11/0.16 |
| | 2 | 2 | 1/5 | 0/0 | 1/1 | 0.05/0.13 |
| | | 1 | 3/12 | 0/0 | 1/1 | 0.16/0.32 |
| 2x2 | 6 | 6 | 0/5 | 0/2 | 0/0.71 | 0/0.06 |
| | | 5 | 0/9 | 0/2 | 0/0.82 | 0/0.12 |
| | 5 | 5 | 0/14 | 0/2 | 0/0.88 | 0/0.28 |
| | | 4 | 0/14 | 0/2 | 0/0.88 | 0/0.28 |
| | | 3 | 0/14 | 0/2 | 0/0.88 | 0/0.28 |
| | 4 | 4 | 3/12 | 7/14 | 0.43/0.46 | 0.1/0.16 |
| | | 3 | 4/19 | 9/16 | 0.44/0.54 | 0.14/0.25 |
| | | 2 | 4/20 | 10/18 | 0.4/0.53 | 0.14/0.26 |
| | 3 | 3 | 8/21 | 17/40 | 0.47/0.34 | 0.28/0.27 |
| | | 2 | 10/35 | 28/62 | 0.36/0.36 | 0.35/0.45 |
| | | 1 | 10/38 | 34/91 | 0.29/0.29 | 0.35/0.49 |
| | 2 | 2 | 10/22 | 66/178 | 0.15/0.11 | 0.34/0.29 |
| | | 1 | 13/41 | 113/305 | 0.12/0.12 | 0.45/0.53 |

Table 2. Performance of ARACNE-MC2 for the small and the large networks. The DPI tolerance is 0.

| | 2x1/3/2 | 2x2/4/3 |
|---|---|---|

| Number of Samples | Precision | Recall | Precision | Recall |
|---|---|---|---|---|
| 1000 | 1 | 0.16 | 0.54 | 0.25 |
| 500 | 1 | 0.13 | 0.58 | 0.25 |
| 250 | 1 | 0.13 | 0.43 | 0.19 |
| 100 | 1 | 0.03 | 0 | 0 |

Table 3: Precision and recall results for selected reaction template / ARACNE supports / surviving ARACNE supports from the large network against a varying number of input metabolic profiles using the ARACNE-MC2 algorithm.